\documentclass[epj]{webofc}\woctitle{CONF12} 
\usepackage[varg]{txfonts}\usepackage{bm}\usepackage{booktabs}
\usepackage{amsmath}\usepackage{color}
\usepackage{dcolumn}\newcolumntype{d}[1]{D{.}{.}{#1}}

\begin{document}\title{Heavy-Meson Decay Constants:\\Isospin
Breaking from QCD Sum Rules}\author{Wolfgang Lucha\inst{1}\fnsep
\thanks{\email{Wolfgang.Lucha@oeaw.ac.at}}\and Dmitri Melikhov
\inst{1,2,3}\fnsep\thanks{\email{dmitri_melikhov@gmx.de}}\and
Silvano Simula\inst{4}\fnsep\thanks{\email{simula@roma3.infn.it}}}
\institute{Institute for High Energy Physics, Austrian Academy of
Sciences, Nikolsdorfergasse 18, A-1050 Vienna, Austria\and
D.~V.~Skobeltsyn Institute of Nuclear Physics, M.~V.~Lomonosov
Moscow State University, 119991, Moscow, Russia\and Faculty of
Physics, University of Vienna, Boltzmanngasse 5, A-1090 Vienna,
Austria\and INFN, Sezione di Roma Tre, Via della Vasca Navale 84,
I-00146 Roma, Italy}

\abstract{Compared with lattice QCD, isospin breaking of
heavy-meson decay constants from QCD sum rules agrees for $D$
mesons but disagrees (by a factor of four) for $B$ mesons.}
\maketitle

\section{Decay constants of heavy--light mesons from QCD sum-rule
perspective}QCD sum rules \cite{SVZ} constitute analytic
relationships between the experimentally measurable properties of
hadrons, on the one hand, and the fundamental parameters of the
underlying quantum field theory of the strong interactions,
quantum chromodynamics (QCD), on the other hand. They therefore
represent a standard (occasionally perhaps indispensable)
theoretical tool of hadron physics. Some time ago, we embarked on
thorough investigations of both the accuracy and precision
actually achievable by and the systematic uncertainties inherent
to the formalism of QCD sum rules
\cite{LMSAU,LMSAUa,LMSAUb,LMSAUc,MAU}. As a consequence of these
studies, we proposed a somewhat modified algorithm that provides
estimates of intrinsic errors
\cite{LMSET,LMSETa,LMSSET,LMSETb,LMSETc}. With this improvement at
our disposal, we revisited QCD sum-rule extractions of a
variety~of~hadronic observables, particularly of the leptonic
decay constants of the heavy--light mesons
\cite{LMSDC,LMSDCa,LMSDCb,LMSDCc,LMSDCd}, which led us to the
conclusion \cite{LMSR,LMSRa,LMSRb,LMSRc,LMSRd} that, in the
bottom-meson system, the decay constants~of the pseudoscalar
mesons exceed those of their vector counterparts, as has been
confirmed~thereafter in lattice QCD \cite{HPQCD}.

For pseudoscalar ($P_q$) and vector ($V_q$) mesons $H_q\equiv
P_q,V_q,$ of mass $M_{H_q},$ built up by a heavy quark $Q=c,b$ and
a light quark $q=u,d$ of masses $m_Q$ and $m_q,$ respectively, the
decay constants,~$f_{H_q},$~of~these mesons [with momentum $p$
and, in the case of vector mesons, polarization vector
$\varepsilon_\mu(p)$] are defined, in terms of suitable,
interpolating heavy--light axial-vector (A) and vector (V)
quark-current operators,~by$$\langle0|\,\bar q\,\gamma_\mu\,
\gamma_5\,Q\,|P_q(p)\rangle={\rm i}\,f_{P_q}\,p_\mu\ ,\qquad
\langle0|\,\bar q\,\gamma_\mu\,Q\,|V_q(p)\rangle=f_{V_q}\,
M_{V_q}\,\varepsilon_\mu(p)\ .$$Our actual aim \cite{LMSIB,LMSIB+}
is the impact of the isospin-spoiling disparity of the $u$- and
$d$-quark masses~\cite{PDG}$$m_u(2\;\mbox{GeV})=
\left(2.3^{+0.7}_{-0.5}\right)\,\mbox{MeV}\ ,\qquad
m_d(2\;\mbox{GeV})=\left(4.8^{+0.5}_{-0.3}\right)\,\mbox{MeV}$$on
the differences $f_{H_d}-f_{H_u}$ of the decay constants of the
charmed and beauty mesons $H=D,D^*,B,B^*.$

\section{Dependence of heavy--light meson decay constants on
light-quark mass}Ignoring electromagnetic and weak interactions,
we focus on the strong interactions of the light quarks $u$ and
$d,$ which then are basically characterized by just their masses
$m_u$ and $m_d,$ respectively. In~order to track the dependence of
the heavy--light meson decay constants on the light-quark masses,
we~consider a fictitious light quark $q$ of mass $m_q$ which we
allow to vary from the average of the $u$ and $d$
masses~\cite{PDG}\begin{equation}m_{ud}\equiv\frac{m_u+m_d}{2}\
,\qquad m_{ud}=\left(3.5^{+0.7}_{-0.2}\right)\,\mbox{MeV}\
,\label{a}\end{equation}up to the mass $m_s$ of the strange quark,
$m_s=(95\pm5)\;\mbox{MeV}$ \cite{PDG}. Such generalization,
however,~enables us to introduce generic decay-constant functions
$f_H(m_q)$ of that continuously varying light-quark mass $m_q,$
derived from our QCD sum-rule approach such that the decay
constants at the $m_q$ values of interest may be defined by the
identification $f_{H_{u|d|ud}}\equiv f_H(m_{u|d|ud}).$ In terms of
the $u$- and $d$-quark mass difference $\delta m\equiv m_d-m_u>0,$
the decay-constant differences $f_{H_d}-f_{H_u}$ thus emerge, at
lowest order in $\delta m,$ from the derivative of $f_H(m_q)$ with
respect to $m_q$ in some vicinity of the $u$ and $d$ masses, say,
at their~average~$m_{ud}$:$$f_{H_d}-f_{H_u}=\left.\frac{\partial
f_H(m_q)}{\partial m_q}\right|_{m_q=m_{ud}}\,\delta m+O(\delta
m^2)\ .$$For simplicity, we prefer to introduce the shifted and
rescaled, dimensionless light-quark mass variable\begin{equation}
x_q\equiv\frac{m_q-m_{ud}}{m_s-m_{ud}}\qquad\Longrightarrow\qquad
\left\{\begin{array}{ll}m_q=m_{ud}&\Longleftrightarrow\ \ \ x_q=0\
,\\m_q=m_s&\Longleftrightarrow\ \ \ x_q=1\ .\end{array}\right.
\label{x}\end{equation}By benefitting from mutual cancellations of
individual errors, a substantial diminution of uncertainties ought
to be gained by discussing, instead of our decay-constant
functions $f_H(x_q)$ themselves, the ratios\begin{equation}
R_H(x_q)\equiv\frac{f_H(x_q)}{f_{H_{ud}}}\label{r}\end{equation}of
the redefined decay-constant functions $f_H(x_q)$ and their values
$f_H(0)\equiv f_{H_{ud}}$ at the mass average $m_{ud}.$ The
derivative of any such decay-constant ratio with respect to $x_q,$
that is, its slope, at the origin~$x_q=0,$ $$R_H^\prime(0)\equiv
\left.\frac{{\rm d}R_H(x_q)}{{\rm d}x_q}\right|_{x_q=0}\ ,$$then
determines the $(\bar Q\,q)$ meson's decay-constant difference
$f_{H_d}-f_{H_u},$ normalized to $f_{H_{ud}},$ according~to
\begin{equation}\frac{f_{H_d}-f_{H_u}}{f_{H_{ud}}}=R_H^\prime(0)\,
\frac{m_d-m_u}{m_s-m_{ud}}\ .\label{d}\end{equation}

\section{Heavy--light meson decay constants: improved QCD sum-rule
formalism}QCD sum rules arise from evaluation of vacuum
expectation values of nonlocal products of convenient
interpolating currents simultaneously at hadronic and QCD levels,
by use of Wilson's operator product expansion for conversion of
nonlocal operators into series of local ones, of Borel
transformations from relevant momenta to Borel variables,
generically called $\tau,$ and of the quark--hadron duality
assumption, asserting equality of hadron-state and
perturbative-QCD contributions beyond effective thresholds $s_{\rm
eff}^{H_q}.$ Taking into account the effective thresholds' Borel
variable dependence $s_{\rm eff}^{H_q}=s_{\rm eff}^{H_q}(\tau)$
proves crucial~for the accuracy of QCD sum-rule outcomes and our
ability to estimate their intrinsic uncertainties \cite{LMSAU,
LMSAUa,LMSAUb,LMSAUc,MAU,LMSET,LMSETa,LMSSET,LMSETb,LMSETc}.
Application of all these procedures to interpolating currents of
type $J={\rm A},{\rm V}$ results in QCD~sum~rules\pagebreak
$$f_{H_q}^2\,M_{H_q}^2\exp\left(-M_{H_q}^2\,\tau\right)
=\int\limits_{(m_Q+m_q)^2}^{s_{\rm eff}^{H_q}(\tau)}{\rm
d}s\exp(-s\,\tau)\,\rho_J(s,m_Q,m_q,m_{\rm sea},\alpha_{\rm
s})+\widehat\Pi_J(\tau,m_Q,m_q,\langle\bar q\,q\rangle,\dots)$$
receiving contributions of purely perturbative origin, which may
be represented by dispersion integrals of spectral densities
$\rho_J(s,m_Q,m_q,m_{\rm sea},\alpha_{\rm s}),$ and
nonperturbative contributions
$\widehat\Pi_J(\tau,m_Q,m_q,\langle\bar q\,q\rangle,\dots)$
parametrized by so-called vacuum condensates (vacuum expectation
values of colour-singlet operators constructed from the degrees of
freedom of QCD) that characterize the properties of the QCD
vacuum.

Picking up the terminology of lattice-QCD practitioners, in the
operator product expansion we take the liberty to discriminate
notationally those quarks which compose the interpolating currents
from~the ``sea'' quarks that contribute only to radiative
corrections and denote all the masses of the~latter by~$m_{\rm
sea}$. As perturbative series expansions in powers of the strong
coupling $\alpha_{\rm s},$ our spectral densities are known
\cite{SD,SDa,JL,G+} fully up to two-loop order $O(\alpha_{\rm s})$
but at three-loop order $O(\alpha_{\rm s}^2)$ only for the case
$m_q=m_{\rm sea}=0.$

Following Ref.~\cite{JL}, we ensure perturbative convergence of
our QCD sum-rule results by adopting for the definition of the
quark masses the modified minimal-subtraction ($\overline{\rm
MS}$) renormalization scheme.

Physical quantities, such as decay constants, cannot depend on
unphysical (renormalization) scales introduced on calculational
grounds. Nevertheless, procedures required by the application of
the QCD sum-rule formalism induce for several reasons artificial
scale dependences of the resulting predictions:\begin{itemize}
\item Unavoidable truncations of perturbative expansions produce
scale dependences of spectral densities.\item Effective thresholds
are determined at given scales such as to reproduce experimental
meson masses.\end{itemize}Although implementation of our advanced
QCD sum-rule algorithms to heavy-meson decay constants removes a
large portion of such scale dependences \cite{LMSDC,LMSDCa,LMSDCb,
LMSDCc,LMSDCd,LMSR,LMSRa,LMSRb,LMSRc,LMSRd}, remaining scale
dependences contribute to the systematic uncertainties of any
extracted QCD sum-rule outcomes. Consequently, in the following we
present our findings for the scales $\mu$ at which the quoted
estimates have been obtained. These~scales are
$\mu=1.7\;\mbox{GeV}$ for the charmed mesons $D$ and $D^*$ and
$\mu=3.75\;\mbox{GeV}$ for the bottom mesons $B$~and~$B^*.$

\begin{table}[b]\centering\caption{Parameter values relevant for
our QCD sum-rule analysis of the decay constants of heavy--light
mesons. }\label{P}\begin{tabular}{lcr}\toprule Parameter in
operator product expansion&Numerical input value&References\\
\midrule$\displaystyle m_{ud}(2\;\mbox{GeV})$&
$(3.70\pm0.17)\;\mbox{MeV}$&\cite{FLAG,FLAGa}\\
$m_s(2\;\mbox{GeV})$&$(93.9\pm1.1)\;\mbox{MeV}$&\cite{FLAG,FLAGa}\\
$m_c(m_c)$&$(1275\pm25)\;\mbox{MeV}$&\cite{PDG}\\
$m_b(m_b)$&$(4247\pm34)\;\mbox{MeV}$&\cite{LMSDCb}\\$\alpha_{\rm
s}(M_Z)$&$0.1184\pm0.0020$& \cite{LMSDCa,LMSDCc,LMSRc}\\
$\displaystyle\langle\bar\ell\,\ell\rangle(2\;\mbox{GeV})\equiv
\frac{\langle\bar u\,u\rangle+\langle\bar d\,d\rangle}{2}$&
$-[(267\pm17)\;\mbox{MeV}]^3$&\cite{LMSDCa,LMSDCc,LMSRc,JL}\\[1.5ex]
$\displaystyle\frac{\langle\bar s\,s\rangle(2\;\mbox{GeV})}
{\langle\bar \ell\,\ell\rangle(2\;\mbox{GeV})}$& $0.8\pm0.3$&
\cite{LMSDCa,LMSDCc,LMSRc,JL}\\[2ex]$\displaystyle
\left\langle\frac{\alpha_{\rm s}}{\pi}\,G\,G\right\rangle$&
$(0.024\pm0.012)\;\mbox{GeV}^4$&\cite{LMSDCa,LMSDCc,LMSRc,JL}\\[1.5ex]
$\displaystyle\frac{\langle\bar\ell\,g_{\rm
s}\,\sigma\,G\,\ell\rangle(2\;\mbox{GeV})}
{\langle\bar\ell\,\ell\rangle(2\;\mbox{GeV})}$&
$(0.8\pm0.2)\;\mbox{GeV}^2$&\cite{LMSDCa,LMSDCc,LMSRc,JL}\\
\bottomrule\end{tabular}\end{table}

The numerical values of the set of parameters employed as input to
our operator product expansion are given in Table~\ref{P}. In
addition, we need an idea about how the mass $M_{H_q}$ of the
fictitious $(\bar Q\,q)$ meson $H_q$ and the vacuum condensate
$\langle\bar q\,q\rangle$ of that fictitious light quark $q$
behave with varying quark mass~$m_q$:\begin{itemize}\item Lattice
QCD \cite{MHa,MHb,MHc,MHd} hints at a linear rise of the masses of
charmed and bottom mesons with $m_q.$ So, for the masses
$M_{H_q}(x_q)$ of the heavy--light ($\bar Q\,q$) mesons, we assume
a linear $x_q$ dependence from the measured \cite{PDG}
nonstrange-meson mass $M_{H_{ud}}$ up to the corresponding
strange-meson mass~$M_{H_s}$~\cite{PDG}:$$M_{H_q}(x_q)=M_{H_{ud}}+
x_q\left(M_{H_s}-M_{H_{ud}}\right)\,.$$\item For the light-quark
condensate $\langle\bar q\,q\rangle,$ we assume a linear $x_q$
dependence from $\langle\bar\ell\,\ell\rangle\equiv (\langle\bar
u\,u\rangle+\langle\bar d\,d\rangle)/2$ at
$m_{ud}\equiv(m_u+m_d)/2$ [cf.~Eq.~(\ref{a})] up to the
strange-quark condensate $\langle\bar s\,s\rangle$ at
$m_s=(95\pm5)\;\mbox{MeV}$~\cite{PDG}:$$\langle\bar
q\,q\rangle=\langle\bar\ell\,\ell\rangle+x_q\left(\langle\bar
s\,s\rangle-\langle\bar\ell\,\ell\rangle\right)\,.$$\end{itemize}

\begin{figure}[b]\centering\begin{tabular}{cc}
\includegraphics[scale=0.33788,clip]{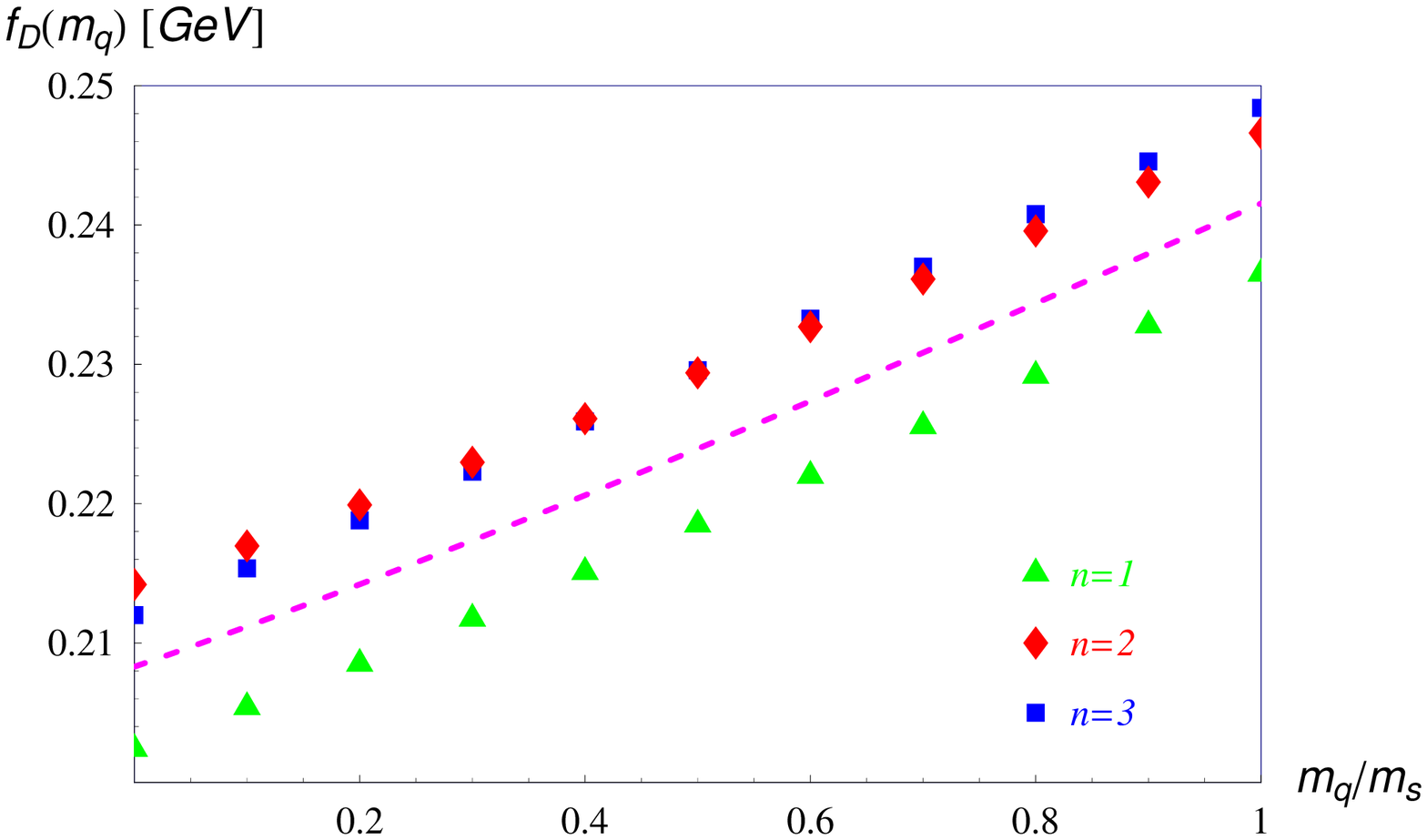}&
\includegraphics[scale=0.33788,clip]{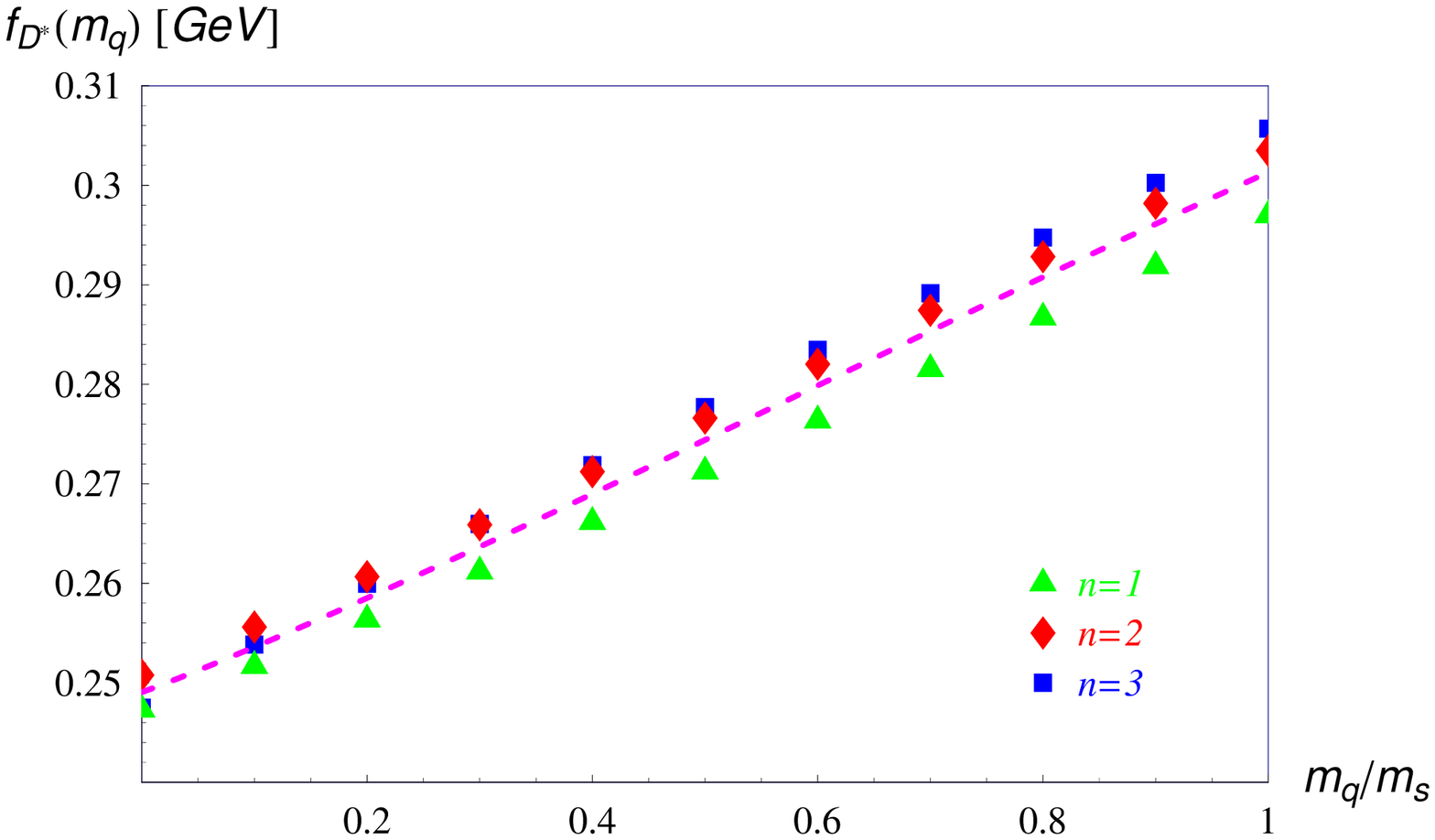}\\(a)&(b)\\[1ex]
\includegraphics[scale=0.33788,clip]{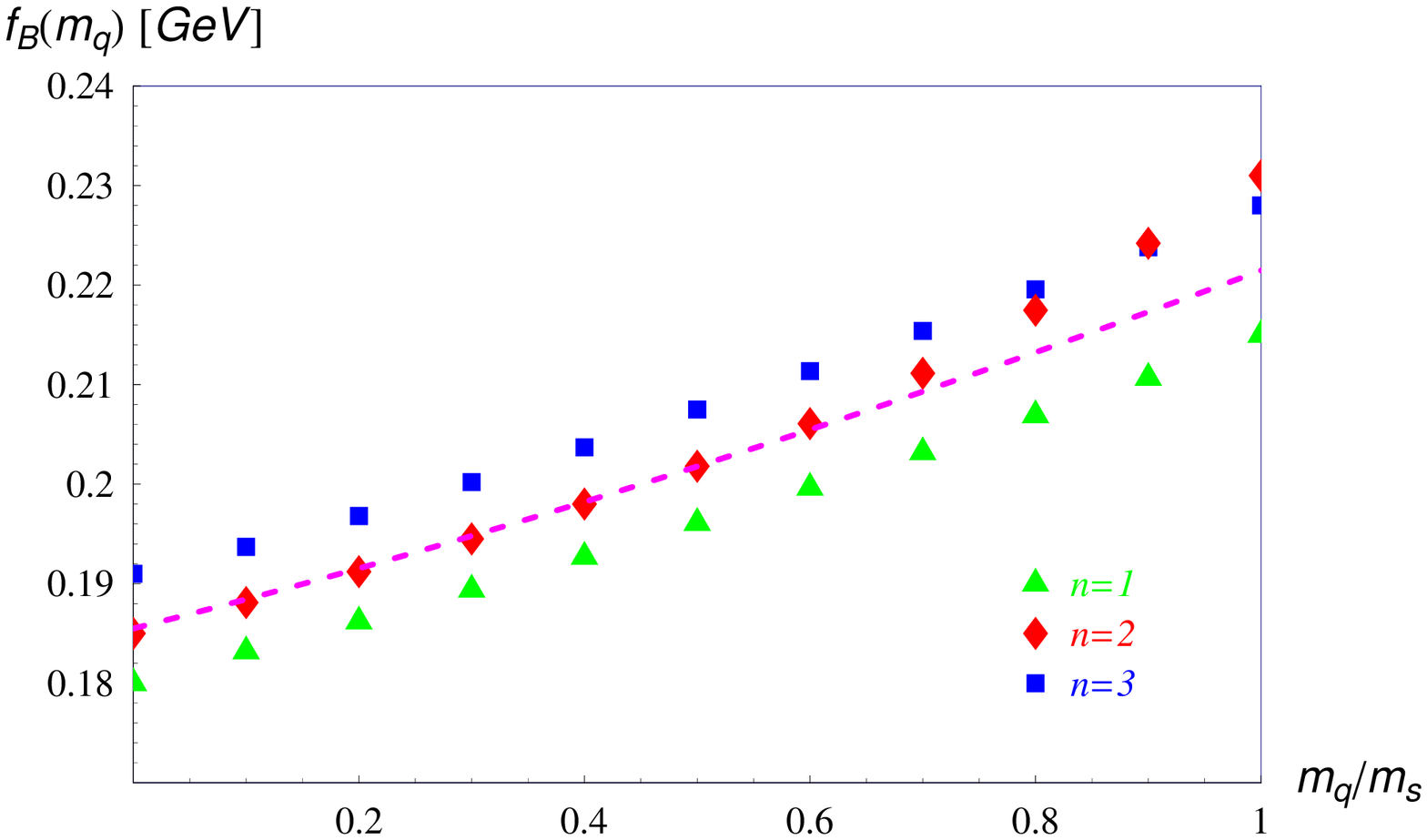}&
\includegraphics[scale=0.33788,clip]{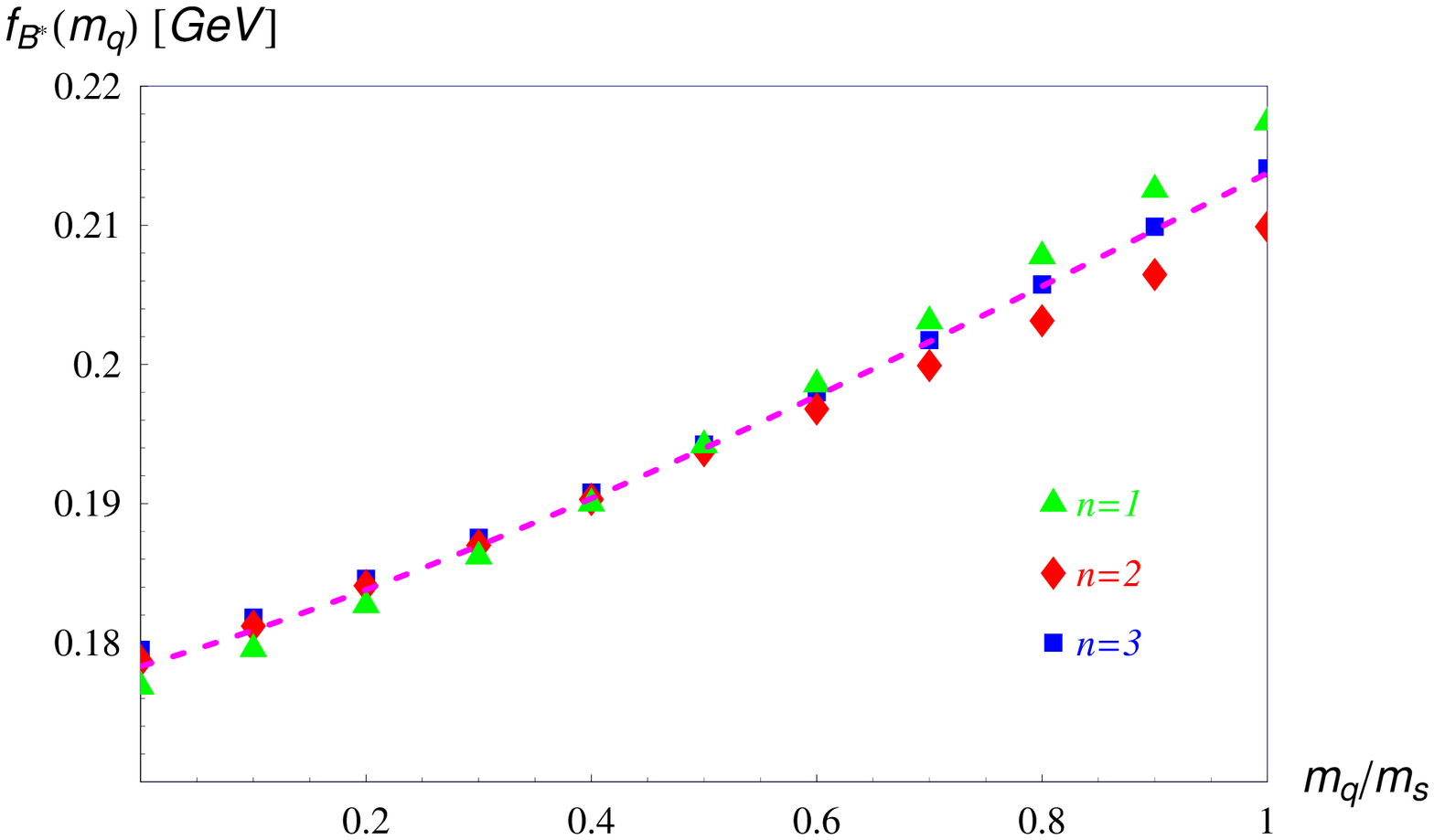}\\(c)&(d)
\end{tabular}\caption{Dependence on the light-quark mass $m_q$
normalized to the strange-quark mass $m_s$ of the decay constants
$f_H(m_q)$ of heavy--light ($\bar Q\,q$) mesons of (a) $D,$ (b)
$D^\ast,$ (c) $B$ and (d) $B^\ast$ type predicted \cite{LMSIB},
with centers indicated by dashed magenta lines, by QCD sum rules
\cite{LMSET,LMSETa,LMSSET,LMSETb,LMSETc} relying on polynomial
\emph{ans\"atze\/} of order $n=1$ (green triangles
\textcolor{green}{$\blacktriangle$}), $n=2$ (red diamonds
\textcolor{red}{$\blacklozenge$}), and $n=3$ (blue squares
\textcolor{blue}{$\blacksquare$}) for the
Borel-parameter-dependent effective thresholds~$s_{\rm
eff}^{H_q}(\tau).$}\label{f}\end{figure}

\begin{figure}[t]\centering\begin{tabular}{cc}
\includegraphics[scale=0.33788,clip]{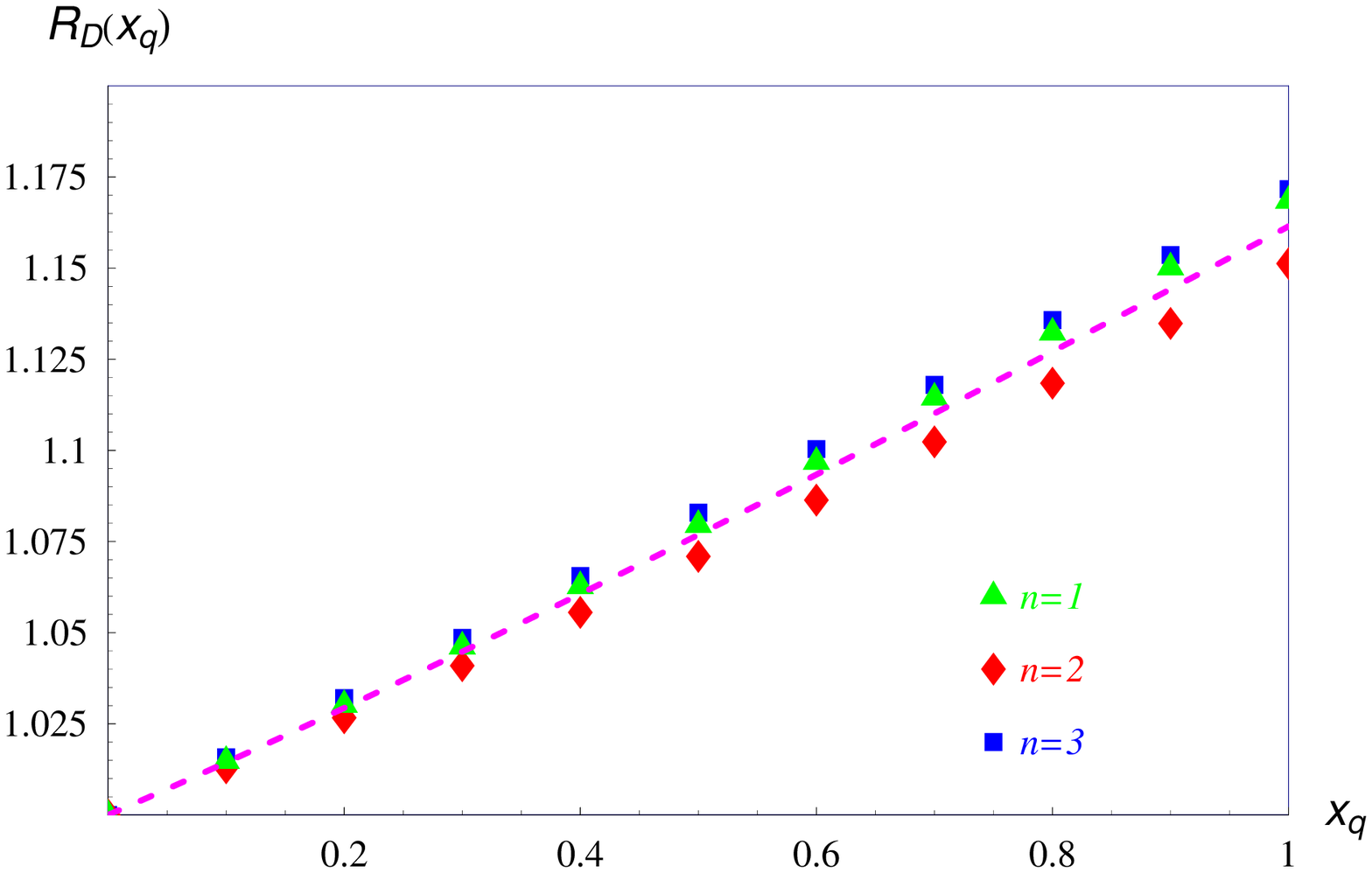}&
\includegraphics[scale=0.33788,clip]{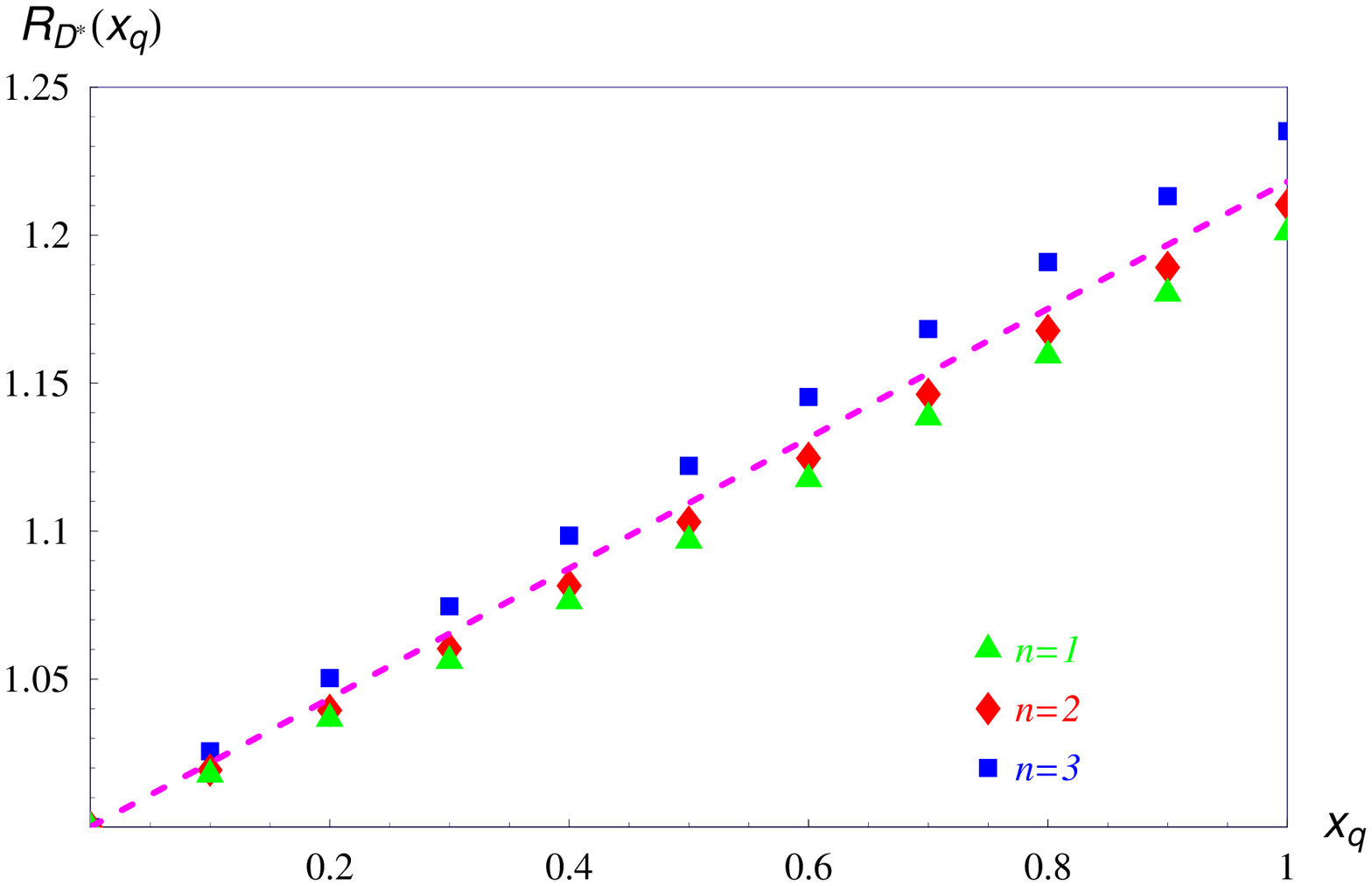}\\(a)&(b)\\[1ex]
\includegraphics[scale=0.33788,clip]{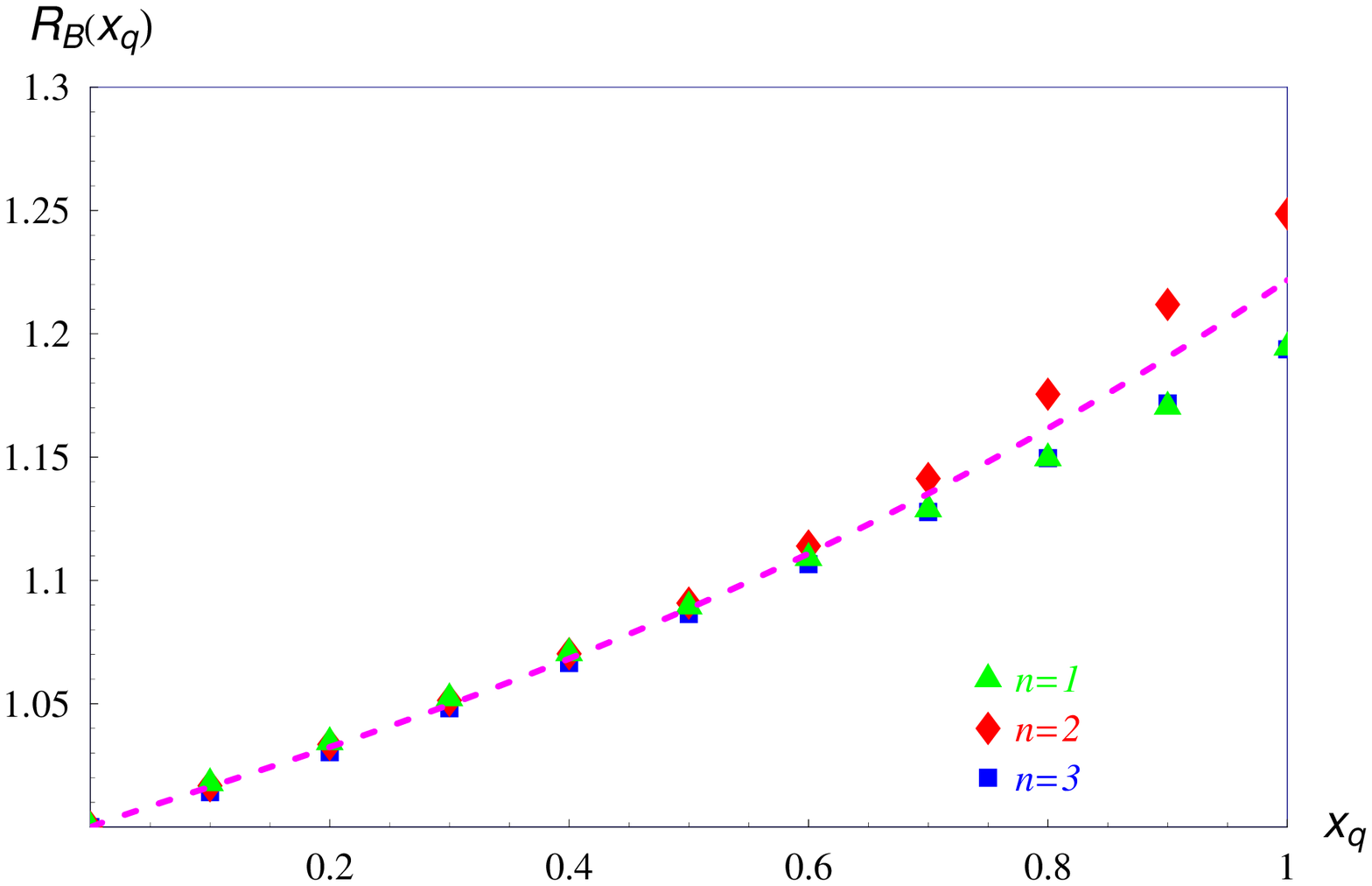}&
\includegraphics[scale=0.33788,clip]{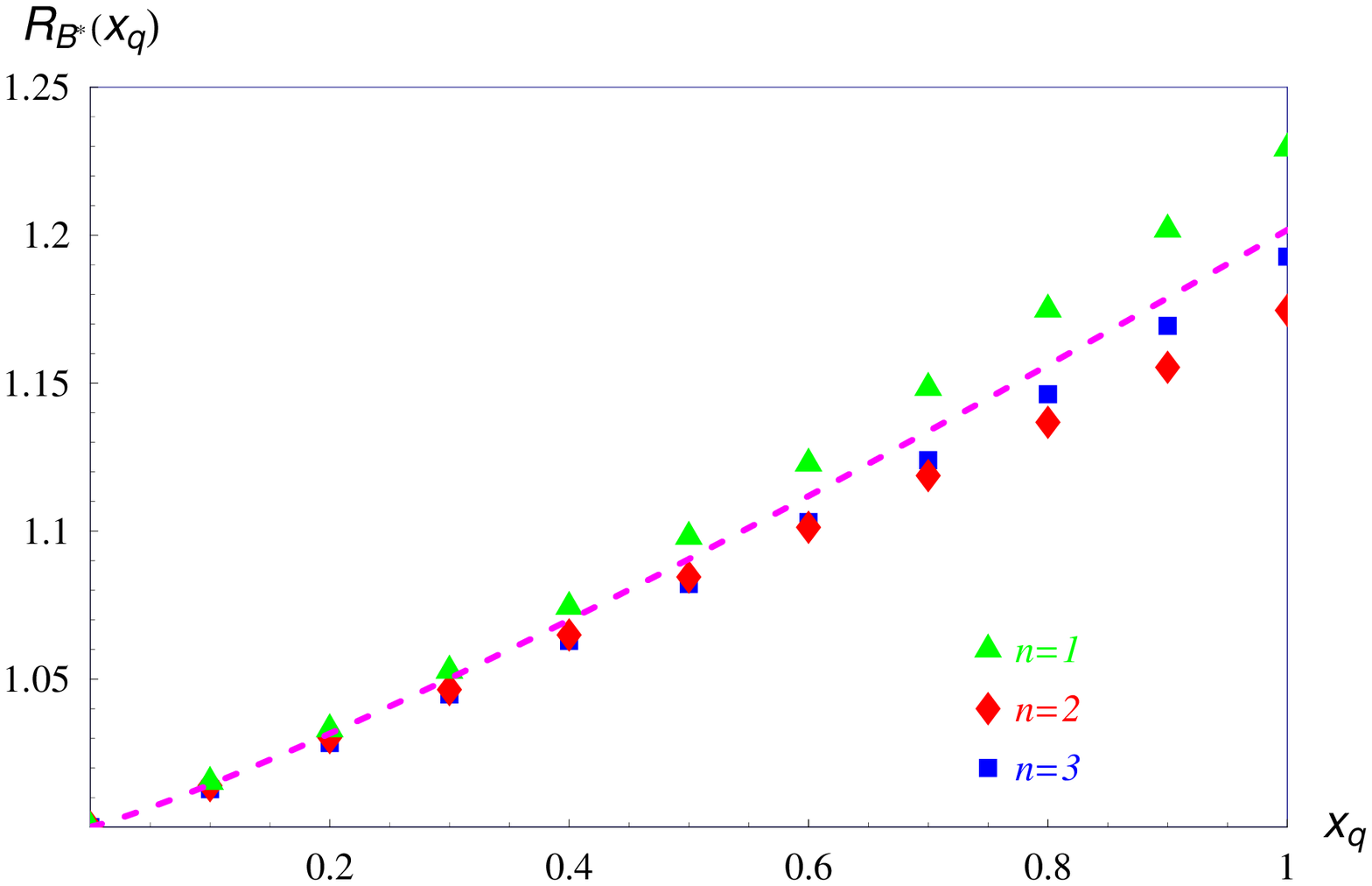}\\(c)&(d)
\end{tabular}\caption{Dependence on the light-quark mass ratio
$x_q$ in Eq.~(\ref{x}) of the ratios $R_H(x_q)\equiv
f_H(x_q)/f_H(0)$ of the~decay constants for heavy--light ($\bar
Q\,q$) mesons of (a) $D,$ (b) $D^\ast,$ (c) $B$ and (d) $B^\ast$
type found \cite{LMSIB+}, with centers indicated by dashed magenta
lines, by QCD sum rules \cite{LMSET,LMSETa,LMSSET,LMSETb,LMSETc}
relying on polynomial \emph{ans\"atze\/} of order $n=1$ (green
triangles \textcolor{green}{$\blacktriangle$}), $n=2$ (red
diamonds \textcolor{red}{$\blacklozenge$}), and $n=3$ (blue
squares \textcolor{blue}{$\blacksquare$}) for the
Borel-parameter-dependent effective thresholds~$s_{\rm
eff}^{H_q}(\tau).$}\label{R}\end{figure}

For the determination of the dependence of the effective
thresholds $s^{H_q}_{\rm eff}(\tau)$ on the Borel variable $\tau,$
it proves to be sufficient to model their behaviour by polynomial
ans\"atze with expansion coefficients~$s^{(n)}_j,$ $$s^{H_q}_{\rm
eff}(\tau)=\sum_{j=0}^ns^{(n)}_j\,\tau^j\ ,$$of rather low order
$n=1,2,3,$ optimized by requiring a satisfactory reproduction of
the experimentally measured meson masses $M_{H_q}$ by the
corresponding QCD sum-rule predictions \cite{LMSET,LMSETa,LMSSET,
LMSETb,LMSETc}. At a given~mass value $m_q,$ the spread of
predictions, obtained by application of our QCD sum-rule
algorithms along~the lines detailed in our earlier decay-constant
extractions \cite{LMSDC,LMSDCa,LMSDCb,LMSDCc,LMSDCd}, for the case
of linear ($n=1$), quadratic ($n=2$), and cubic ($n=3$) behaviour
of the effective thresholds $s^{H_q}_{\rm eff}(\tau)$ provides, by
its central value~and half-width, estimates of the decay-constant
functions $f_H(m_q)$ and ratios $R_H(x_q),$ Eq.~(\ref{r}), as well
as their systematic uncertainties
\cite{LMSET,LMSETa,LMSSET,LMSETb,LMSETc}, shown for the mesons
$H_q=D,D^*,B,B^*$ in Figs.~\ref{f} and \ref{R}, respectively.

\section{Decay-constant differences from the slopes of
decay-constant functions}In order to eventually extract from the
QCD sum-rule outcomes in Fig.~\ref{R} the slopes $R_H^\prime(0)$
determining, as expressed by Eq.~(\ref{d}), the decay-constant
differences $f_{H_d}-f_{H_u},$ we describe the behaviour of any
ratio $R_H(x_q)$ as function of the mass-ratio variable $x_q$ in
its respective $x_q$ interval, in three different ways, two of
na\"ive polynomial shape and one inspired by heavy-meson chiral
perturbation theory~(HM$\chi$PT)~\cite{SZ}:\begin{subequations}
\begin{align}R_H(x_q)&=1+a^H\,x_q\ ,&&0<x_q<0.4\qquad\mbox{(linear
ansatz)}\ ,\label{l}\\R_H(x_q)&=1+b_1^H\,x_q+b_2^H\,x_q^2\ ,&&
0<x_q<1\qquad\mbox{(quadratic ansatz)}\label{q}\ ,\\R_H(x_q)&=
1+R_\chi(m_q,m_{ud},m_s)+c_1^H\,x_q+c_2^H\,x_q^2\ ,&&0<x_q<1\qquad
\mbox{(HM$\chi$PT ansatz)}\label{pq}\ .\end{align}
\end{subequations}In the HM$\chi$PT case, the term
$R_\chi(m_q,m_{ud},m_s)$ is independent of the mesons $H_q$ under
consideration;~its (somewhat lengthy) explicit expression can be
found in the Appendix of Ref.~\cite{LMSIB+}. The parameters~$a^H,$
$b_i^H$ and $c_i^H$ ($i=1,2$) may be utilized to optimize our
fits. Table~\ref{s} summarizes our findings for the~slopes
$R_H^\prime(0),$ derived by fitting the $x_q$ dependences of the
ratios $R_H(x_q)$ depicted in Fig.~\ref{R} for each of the~three
ans\"atze (\ref{l}), (\ref{q}), and (\ref{pq}), as well as the
averages of the three individual results, for $H=D,D^*,B,B^*.$

\begin{table}[hbt]\centering\caption{Numerical values of the
decay-constant-governing slope $R'_H (0)$ of the light-quark-mass
dependent ratio $R_H(x_q)$ at the average $u$-$d$ mass point
$x_q=0,$ resulting from the three fits given by Eqs.~(\ref{l}),
(\ref{q}) and (\ref{pq}) in their respective $x_q$ intervals, and
of the corresponding averages for the charmed and bottom mesons
$H=D,D^*,B,B^*.$}\begin{tabular}{ld{1.10}d{1.10}d{1.10}d{1.10}}
\toprule Meson&\multicolumn{4}{c}{$R'_H(0)$}\\&\multicolumn{1}{c}
{linear fit (\ref{l})}&\multicolumn{1}{c}{quadratic fit
(\ref{q})}&\multicolumn{1}{c}{HM$\chi$PT fit
(\ref{pq})}&\multicolumn{1}{c}{average}\\
&\multicolumn{1}{c}{$0<x_q<0.4$}&\multicolumn{1}{c}{$0<x_q<1$}
&\multicolumn{1}{c}{$0<x_q<1$}&\\\midrule
$D$&0.148\pm0.007&0.144\pm0.009&0.171\pm0.009&0.154\pm0.015\\
$D^*$&0.218\pm0.016&0.218\pm0.021&0.248\pm0.022&0.228\pm0.024\\[.5ex]
$B$&0.168\pm0.004&0.146\pm0.008&0.174\pm0.008&0.163\pm0.014\\
$B^*$&0.156\pm0.006&0.139\pm0.009&0.162\pm0.010&0.152\pm0.013\\
\bottomrule\end{tabular}\label{s}\end{table}

For the lattice-QCD value $\delta m(2\;\mbox{GeV})=(2.67\pm0.22)\;
\mbox{MeV}$ \cite{FLAG,FLAGa} of the $u$-$d$ mass difference and
the averages of Table~\ref{s}, Eq.~(\ref{d}) predicts, for the
normalized decay-constant differences
$(f_{H_d}-f_{H_u})/f_{H_{ud}},$\begin{align*}
\frac{f_{D^\pm}-f_{D^0}}{f_D}&=0.0046(6)\ ,&
\frac{f_{D^{*\pm}}-f_{D^{*0}}}{f_{D^*}}&=0.0067(9)\ ,\\
\frac{f_{B^0}-f_{B^\pm}}{f_B}&=0.0048(6)\ ,&
\frac{f_{B^{*0}}-f_{B^{*\pm}}}{f_{B^*}}&=0.0045(5)\ ,\end{align*}
indicating that the decay constants satisfy the inequality
$f_{H_d}>f_{H_u}$ for all the mesons $H=D,D^*,B,B^*.$
Multiplication of the above ratios by the decay constants
collected in Table~\ref{fH} entails, as our final results for the
decay-constant differences of charmed and bottom heavy--light
mesons due to isospin~breaking,\begin{align*}
f_{D^\pm}-f_{D^0}&=(0.95\pm0.13)\;\mbox{MeV}\ ,&
f_{D^{*\pm}}-f_{D^{*0}}&=(1.69\pm0.27)\;\mbox{MeV}\ ,\\
f_{B^0}-f_{B^\pm}&=(0.92\pm0.13)\;\mbox{MeV}\ ,&
f_{B^{*0}}-f_{B^{*\pm}}&=(0.82\pm0.11)\;\mbox{MeV}\ .\end{align*}

\begin{table}[h]\centering\caption{Decay constants $f_H$ of the
charmed and beauty mesons $H=D,D^*,B,B^*$: numerical values
\cite{LMSDCa,LMSDCb,LMSDCc,LMSRc}.}\begin{tabular}{ld{1.7}r}
\toprule Meson $H$&\multicolumn{1}{c}{Decay constant $f_H$\;(MeV)}
&Reference\\\midrule
$D$&206.2\pm8.9&\cite{LMSDCa}\\$D^*$&252.2\pm22.7&\cite{LMSDCc}\\[.5ex]
$B$&192.0\pm14.6&\cite{LMSDCb}\\$B^*$&181.8\pm13.7&\cite{LMSRc}\\
\bottomrule\end{tabular}\label{fH}\end{table}

Table~\ref{C} confronts our predictions \cite{LMSIB+} with
available lattice-QCD outcomes for the decay-constant differences
$f_{H_d}-f_{H_u}$ of \emph{pseudoscalar\/} heavy--light mesons: In
the case of the $D$ mesons, we find~a~really nice agreement. In
the case of the $B$ mesons, however, our QCD sum-rule predictions
are by a factor of four smaller than respective lattice-QCD
claims, betraying a tension of some three standard deviations.
Confidence in our approach may be drawn from the fact that our
results are of very similar size as those reported by lattice QCD
for the $K$ mesons \cite{LIBK}: both systems lead to differences
of the order~of~$1\;\mbox{MeV}.$

\begin{table}[ht]\centering\caption{Decay-constant difference of
\emph{pseudoscalar\/} heavy mesons: QCD sum rule \cite{LMSIB+}
vs.\ lattice QCD \cite{LIBb,LIB+,LIBc,PDGIB}.}\label{C}
\begin{tabular}{ld{1.8}d{1.6}r}\toprule Decay-constant difference&
\multicolumn{1}{c}{QCD sum rule \cite{LMSIB+}}&\multicolumn{1}{c}
{Lattice QCD}&\multicolumn{1}{c}{References}\\\midrule
$f_{D^\pm}-f_{D^0}\;\mbox{(MeV)}$&0.95\pm0.13&
0.94_{-0.12}^{+0.50}&\cite{LIBb}\\
$f_{B^0}-f_{B^\pm}\;\mbox{(MeV)}$&0.92\pm0.13&3.8\pm1.0&
\cite{LIB+,LIBc,PDGIB}\\\bottomrule\end{tabular}\end{table}

\acknowledgement{D.~M.\ would like to express gratitude for
support by the Austrian Science Fund (FWF) under project
P29028-N27.}


\begin{thebibliography}{99}
\bibitem{SVZ}M.~A.~Shifman, A.~I.~Vainshtein, and V.~I.~Zakharov,
Nucl.~Phys.~B {\bf 147} (1979) 385.
\bibitem{LMSAU}W.~Lucha, D.~Melikhov, and S.~Simula, Phys.~Rev.~D
{\bf 76} (2007) 036002, arXiv:0705.0470 [hep-ph].
\bibitem{LMSAUa}W.~Lucha, D.~Melikhov, and S.~Simula, Phys.~Lett.~B
{\bf 657} (2007) 148, arXiv:0709.1584 [hep-ph].
\bibitem{LMSAUb}W.~Lucha, D.~Melikhov, and S.~Simula,
Phys.~At.~Nucl.~{\bf 71} (2008) 1461.
\bibitem{LMSAUc}W.~Lucha, D.~Melikhov, and S.~Simula, Phys.~Lett.~B
{\bf 671} (2009) 445, arXiv:0810.1920 [hep-ph].
\bibitem{MAU} D.~Melikhov, Phys.~Lett.~B {\bf 671} (2009) 450,
arXiv:0810.4497 [hep-ph].
\bibitem{LMSET}W.~Lucha, D.~Melikhov, and S.~Simula, Phys.~Rev.~D
{\bf 79} (2009) 096011, arXiv:0902.4202 [hep-ph].
\bibitem{LMSETa}W.~Lucha, D.~Melikhov, and S.~Simula, J.~Phys.~G
{\bf 37} (2010) 035003, arXiv:0905.0963 [hep-ph].
\bibitem{LMSSET} W.~Lucha, D.\ Melikhov, H.~Sazdjian, and S.~Simula,
Phys.~Rev.~D {\bf 80} (2009) 114028, arXiv: 0910.3164 [hep-ph].
\bibitem{LMSETb}W.~Lucha, D.~Melikhov, and S.~Simula, Phys.~Lett.~B
{\bf 687} (2010) 48, arXiv:0912.5017 [hep-ph].
\bibitem{LMSETc}W.~Lucha, D.~Melikhov, and S.~Simula,
Phys.~At.~Nucl.~{\bf 73} (2010) 1770, arXiv:1003.1463 [hep-ph].
\bibitem{LMSDC}W.~Lucha, D.~Melikhov, and S.~Simula, J.~Phys.~G
{\bf 38} (2011) 105002, arXiv:1008.2698 [hep-ph].
\bibitem{LMSDCa}W.~Lucha, D.~Melikhov, and S.~Simula, Phys.~Lett.~B
{\bf 701} (2011) 82, arXiv:1101.5986 [hep-ph].
\bibitem{LMSDCb}W.~Lucha, D.~Melikhov, and S.~Simula, Phys.~Rev.~D
{\bf 88} (2013) 056011, arXiv:1305.7099 [hep-ph].
\bibitem{LMSDCc}W.~Lucha, D.~Melikhov, and S.~Simula, Phys.~Lett.~B
{\bf 735} (2014) 12, arXiv:1404.0293 [hep-ph].
\bibitem{LMSDCd}W.~Lucha, D.~Melikhov, and S.~Simula,
EPJ Web Conf.~{\bf 80} (2014) 00043, arXiv:1407.5512 [hep-ph].
\bibitem{LMSR}W.~Lucha, D.~Melikhov, and S.~Simula, EPJ Web
Conf.~{\bf 80} (2014) 00046, arXiv:1410.6684~[hep-ph].
\bibitem{LMSRa}W.~Lucha, D.~Melikhov, and S.~Simula,
arXiv:1411.3890 [hep-ph].
\bibitem{LMSRb}W.~Lucha, D.~Melikhov, and S.~Simula, AIP
Conf.~Proc.~{\bf 1701} (2016) 050007, arXiv:1411.7844 [hep-ph].
\bibitem{LMSRc}W.~Lucha, D.~Melikhov, and S.~Simula, Phys.~Rev.~D
{\bf 91} (2015) 116009, arXiv:1504.03017 [hep-ph].
\bibitem{LMSRd}W.~Lucha, D.~Melikhov, and S.~Simula, PoS
(EPS-HEP2015) 532, arXiv:1508.07595 [hep-ph].
\bibitem{HPQCD}B.~Colquhoun \emph{et al.}, HPQCD Coll.,
Phys.~Rev.~D {\bf 91} (2015) 114509, arXiv:1503.05762 [hep-lat].
\bibitem{LMSIB}W.~Lucha, D.~Melikhov, and S.~Simula, HEPHY-PUB
971/16 (2016), arXiv:1609.02382 [hep-ph].
\bibitem{LMSIB+}W.~Lucha, D.~Melikhov, and S.~Simula, HEPHY-PUB
973/16 (2016), RM3-TH/16-9, arXiv: 1609.05050 [hep-ph].
\bibitem{PDG}K.~A.~Olive \emph{et al.}, Particle Data Group,
Chin.~Phys.~C {\bf 38} (2014) 090001.
\bibitem{SD}K.~G.~Chetyrkin and M.~Steinhauser, Phys.~Lett.~B {\bf
502} (2001) 104, arXiv:hep-ph/0012002.
\bibitem{SDa}K.~G.~Chetyrkin and M.~Steinhauser, Eur.~Phys.~J.~C
{\bf 21} (2001)~319, arXiv:hep-ph/0108017.
\bibitem{JL}M.~Jamin and B.~O.~Lange, Phys.~Rev.~D {\bf 65} (2002)
056005, arXiv:hep-ph/0108135.
\bibitem{G+}P.~Gelhausen, A.~Khodjamirian, A.~A.~Pivovarov, and
D.~Rosenthal, Phys.~Rev.~D {\bf 88} (2013) 014015, arXiv:1305.5432
[hep-ph]; {\bf 89} (2014) 099901(E); {\bf 91} (2015) 099901(E).
\bibitem{FLAG}S.~Aoki \emph{et al.}, FLAG Working Group,
Eur.~Phys.~J.~C {\bf 74} (2014) 2890, arXiv:1310.8555 [hep-lat].
\bibitem{FLAGa}S.~Aoki \emph{et al.}, FLAG Working Group,
arXiv:1607.00299 [hep-lat].
\bibitem{MHa}B.~Blossier \emph{et al.}, Phys.~Rev.~D {\bf 82}
(2010) 114513, arXiv:1010.3659 [hep-lat].
\bibitem{MHb}P.~Dimopoulos \emph{et al.}, ETM Coll., J.~High Energy
Phys.~01 (2012) 046, arXiv:1107.1441 [hep-lat].
\bibitem{MHc}N.~Carrasco \emph{et al.}, ETM Coll., Nucl.~Phys.~B
{\bf 887} (2014) 19, arXiv:1403.4504 [hep-lat].
\bibitem{MHd}A.~Bussone \emph{et al.}, ETM Coll., Phys.~Rev.~D {\bf
93} (2016) 114505, arXiv:1603.04306 [hep-lat].
\bibitem{SZ}S.~R.~Sharpe and Y.~Zhang, Phys.~Rev.~D {\bf 53} (1996)
5125, arXiv:hep-lat/9510037.
\bibitem{LIBb}A.~Bazavov \emph{et al.}, Fermilab Lattice and MILC
Collaborations, Phys.~Rev.~D {\bf 90} (2014) 074509,
arXiv:1407.3772 [hep-lat].
\bibitem{LIB+}R.~J.~Dowdall \emph{et al.}, HPQCD Coll.,
Phys.~Rev.~Lett.~{\bf 110} (2013) 222003, arXiv:1302.2644
[hep-lat].
\bibitem{LIBc}N.~H.~Christ \emph{et al.}, RBC and UKQCD
Collaborations, Phys.~Rev.~D {\bf 91} (2015) 054502,
arXiv:1404.4670 [hep-lat].
\bibitem{PDGIB}J. L.~Rosner, S.~Stone, and R.~S.~Van der Water,
arXiv:1509.02220 [hep-ph].
\bibitem{LIBK}N.~Carrasco \emph{et al.}, ETM Coll., Phys.~Rev.~D
{\bf 91} (2015) 054507, arXiv:1411.7908 [hep-lat].
\end{thebibliography}
\end{document}